\documentstyle[amssymb,prl,aps,epsfig,twocolumn]{revtex}
\begin{document}

\draft

\twocolumn[\hsize\textwidth\columnwidth\hsize\csname@twocolumnfalse\endcsname
\title{Two-Qubits Entanglement Induced by a Faster Data Bus}
\author{P. Zhang, Y. D. Wang and C. P. Sun $^{a,b}$}
\address{Institute of Theoretical Physics, the Chinese Academy of Science,\\
Beijing, 100080, China}
\maketitle
\begin{abstract}
We propose a theoretical protocol to create the entanglement of two qubits via the
Born-Oppenheimer (BO) approximation. In our scheme, each qubit is coupled to
a faster data bus whose frequency is much larger than the energy
spacing of the qubits and thus the BO approximation is valid. Then the
adiabatic separation of qubits from the data bus can induce an effective
potential to couple the two qubits, which can be utilized to create a
quantum logic gate. We also discuss the quantum decoherence caused by the
adiabatic entanglement between the two qubits and the external field.
\end{abstract}
 ]

\section{Introduction}

In quantum information theory, it is important to create a "maximum
entanglement" between two qubits. It has been shown by Barenco {\it et. al.} 
\cite{Gate2} that, together with single-bit operations, the entanglement
based two qubit gates except the classical SWAP gate (including, the CNOT 
\cite{Gate} and controlled phase gates) forms a universal set of logic gates
of quantum computation. In other words, any unitary transformation of single
or many-qubit system can be decomposed into the quantum networks of
single-bit gates and non-trivial two-qubit gates . In the last few years,
many schemes to create two-qubit interaction have been suggested and
experimentally implemented [3-8]. In this paper, we will propose a new
scenario to induce a kind of two qubit interaction creating efficient
quantum entanglement. In our proposal, each qubit interacts with a common
quantum field as a data bus. If the frequency of the quantum field is much
larger than the energy spacing of the qubit, the degree of freedom of the
quantum field and the variables of the two qubit system can be separated
adiabatically. In this situation, an effective two-qubit interaction can be
induced by means of the Born-Oppenheimer (BO) approximation. To explain the
main idea in our proposal in details, we will first briefly review the
relevant quantum adiabatic theorem and the BO approximation as follows.

Adiabatic approximation \cite{aa} is widely used in many fields of physics
including classical physics and quantum physics. In the latter, one has a
rich applications of quantum adiabatic approximation (QAA). We consider a
time-dependent quantum system. Its Hamiltonian $H\left( t\right) =H\left[
R_{i}\left( t\right) \right] $ is controlled by some adiabatically changing
parameters $R_{i}\left( t\right) $s. If the system is initially prepared in
the $n$-th instantaneous eigen-state of the Hamiltonian $H\left( 0\right) $
at $t=0$, then during the evolution the quantum system will still persist in
the $n$-th eigen-state of the instantaneous Hamiltonian $H\left( t\right) $
at another instance. The evolution of the system under the control of $%
R_{i}\left( t\right) $s can be evaluated by QAA, but the back action of the
quantum system on the controlling parameters $R_{i}\left( t\right) $ is
neglected. Of course, the QAA can only work well when $R_{i}\left( t\right) $
behave classically.

When the Hamiltonian of the quantum system depends on the variables of
another slowly varying quantum system, (e.g., the motion of an electron near
nuclear), one needs a more exact approximation, in which the interactions
between the fast system and the slow one, especially for the "back action"
of the fast one on the slow one, are considered. To this end, in 1930 \cite%
{BO}, Born and Oppenheimer suggested a high order approximation (we now call
it Born-Oppenheimer (BO) approximation) to tackle this problem. In the view
of the BO approximation, to solve the dynamics of a composite system with
fast and slowly changing parts, we can first solve the Schr$\ddot{o}$dinger
equation of the fast part for each fixed slow variable $q$. Then the
obtained $q$-dependent eigen-value $V_{n}(q)$ provides the slow part with an
effective potential when the motion of the slow part can not excite the
transitions among those fast states, i.e., the adiabatic condition holds. In
this sense, the time dependent wave function of the composite system can be
written as 
\begin{equation}
\Psi \left( q,x;t\right) =\psi _{n}\left( q,x\right) \phi \left( q,t\right) .
\end{equation}%
Here, $x$ is the variable of the fast part, $\psi _{n}\left( q,x\right) $
the $q$-dependent eigen-state of the fast part corresponding to eigenvalue $%
V_{n}(q);$ and the slow part $\phi \left( q,t\right) $ evolves governed by
the effective Hamiltonian $\hat{H}_{f}+V_{n}\left( \hat{q}\right) ,$ i.e. $%
\phi \left( q,t\right) =e^{-i\left[ \hat{H}_{s}+V_{n}\left( \hat{q}\right) %
\right] t}\phi \left( q,0\right) ,$ where $\hat{H}_{s}$ is the free
Hamiltonian of the slow part.

In this paper we make a crucial observation for the creation of the
nontrivial two qubits interaction of logical gate in quantum information as
a result from the application of the BO approximaiton to a three body
system. According to above discussion, it is very interesting that the slow
part can be divided into two subsystems with variables $q_{1}$\ and $q_{2}$\
without interaction between them. In this situation, the Hamiltonian of the
composite system can be generally written as%
\begin{equation}
H=H_{1}\left( q_{1}\right) +H_{2}\left( q_{2}\right) +W_{1}\left(
q_{1},x\right) +W_{2}\left( q_{2},x\right) +H_{f}\left( x\right)  \label{a1}
\end{equation}%
where $H_{1}\left( q_{1}\right) $, $H_{2}\left( q_{2}\right) $ and $%
H_{f}\left( x\right) $ are the free Hamiltonians of the two slow subsystems
( we can note them as{\bf \ }$q_{1},q_{2}$ ) and the fast subsystem. $%
W_{1}\left( q_{1},x\right) $ ($W_{2}\left( q_{2},x\right) $) are the
interaction Hamiltonians between $q_{1}$ ($q_{2}$) and $x$. As we have
emphasized, Eq. (\ref{a1}) implies that there is no direct interaction
between $q_{1}$ and $q_{2}$. Under adiabatic condition, the eigenvalue $%
V_{n}(q_{1},q_{2})$ of the Hamiltonian $W_{1}\left( q_{1},x\right)
+W_{2}\left( q_{2},x\right) +H_{f}\left( x\right) ${\bf \ }of fast part
induces an effective potential on the slow variables $q_{1}$ and $q_{2}$. In
usual case, $V_{n}(q_{1},q_{2})$ can not be decomposed into the sum of two
independent potentials of $q_{1}$ and $q_{2}$ i.e. $V_{n}(q_{1},q_{2})\neq
V_{1n}(q_{1})+V_{2n}(q_{2})$. Therefore, an effective interaction between $%
q_{1}$ and $q_{2}$ is induced via the BO approximation and then one can
regard the fast variable $x$\ as a data bus which can be removed. Indeed, in
quantum information process, the inter-qubit interaction is very crucial to
create the nontrivial interaction between two qubits $q_{1}$ and $q_{2}$ if
we can regard the above mentioned two subsystems of $q_{1}$ and $q_{2}$ as
two qubits.

We should point out that, in many previous scenarios of quantum logical gate%
{\bf ,} an external field was usually used as a data bus where the quantum
information carried by a qubit is stored and then transferred to another
qubit. Most of those proposals rely on the rotation wave approximation
(RWA). The effective interaction between the two qubit system can be
obtained through the adiabatic elimination of the external field \cite%
{adiabatic elimination}.

However, to adiabatically eliminate the degree of freedom of the external
field, the coupling strength $g$ between the qubit and the external field
ought to be weak enough so that the ratio $\frac{g}{\Delta }$\ is very
small. Here, $\Delta $ is the detuning i.e. $\Delta =\left\vert \omega
_{a}-\omega \right\vert $ where $\omega _{a}$ is the transition frequency of
the qubit and $\omega $ is the frequency of the external field. On the other
hand, since the RWA requires $\Delta \leq \omega _{a},$ the above large
detuning condition implies that $g<<\omega _{a}$. Then the effective
coupling strength between the two qubit system $\frac{g^{2}}{\Delta }$ is
much smaller than $\omega _{a}$. This weak coupling may reduce the
\textquotedblright quality factor\textquotedblright\ $Q$ of the logic gate.
Here $Q$ can be understood as the ratio between the atomic life time{\bf \ }%
and the logic operation time{\bf \ }$\tau _{l}\sim \left( \frac{g^{2}}{%
\Delta }\right) ^{-1}$ since the Rabi transition frequency can be defined as
an upper limit of the operation frequency.

In our proposal based on the BO approximaiton , the frequency of the
external field can be significantly larger than the energy spacing of the
qubits. Then $g$ may be larger than $\omega _{a}$ or, at least, have the
value in the same order as $\omega _{a}$. The ratio between effective
interaction strength $\frac{g^{2}}{\Delta }$ and $\omega _{a}$ can be much
stronger than that in the previous scenarios mentioned above. In the next
section, there is a brief introduction of our quantum logic gate model. The
precise analysis of our model based on the perturbation theory is given in
section III. In section IV, the adiabatic decoherence effect is analyzed
detailedly in association with the idea of the adiabatic entanglements
presented recently by Sun et. al. \cite{sun1}. We also obtain the condition
under which the decoherence effect can be neglected and the two-qubit
interaction model works well. There are some conclusions and discussions
about the physical realization of our{\bf \ }protocol in the last section.

\section{Effective Interaction between Two Qubits}

We generally consider a system consisting of two qubits with the same energy
spacing $\omega _{a}$ and an external field which is described as a dynamic
data bus. We can treat the external field as a harmonic oscillator with
frequency $\omega $ (see Fig. 1).

\begin{figure}[h]
\begin{center}
\includegraphics[width=6cm,height=3cm]{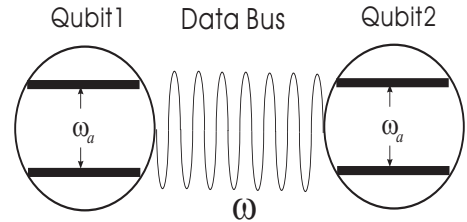} \vspace{0.3cm}
\end{center}
\caption{ The interaction between two qubits and a external field. When $%
\protect\omega >>\protect\omega _{a}$, the external works as a faster data
bus}
\end{figure}

The Hamiltonian of the composite system reads 
\begin{equation}
H=\omega _{a}\left( \hat{\sigma}_{z}^{\left( 1\right) }+\hat{\sigma}%
_{z}^{\left( 2\right) }\right) +g\left( \hat{a}^{\dagger }+\hat{a}\right)
\left( \hat{\sigma}_{x}^{\left( 1\right) }+\hat{\sigma}_{x}^{\left( 2\right)
}\right) +\omega \hat{a}^{\dagger }\hat{a}  \label{a}
\end{equation}%
where $\hat{\sigma}_{z}^{\left( i\right) }=\left\vert e\right\rangle
_{i}\left\langle e\right\vert -\left\vert g\right\rangle _{i}\left\langle
g\right\vert $, $\hat{\sigma}_{x}^{\left( i\right) }=\left\vert
e\right\rangle _{i}\left\langle g\right\vert +\left\vert g\right\rangle
_{i}\left\langle e\right\vert ,\left\vert e\right\rangle _{i}$ and $%
\left\vert g\right\rangle _{i}$ are the excited and ground states of the $i$%
-th qubit; $\hat{a}^{\dagger }$ and $\hat{a}$ the creation and annihilation
operator of the external field, and $g$ the coupling strength between the
qubits and this field. This model system can be realized in many practical
physical situations such as the two trapped ion with the spatial oscillation
mode, two atoms in a single mode cavity and the two Josephson charge qubits
coupling to a large Josephson or{\bf \ }nonamechanical resonator \cite{nano}
. We assume $\omega $ is much larger than $\omega _{a}$. In this case, the
RWA could not be applied since the conditon for the validity of RWA is $%
\left\vert \omega +\omega _{a}\right\vert >>\left\vert \omega -\omega
_{a}\right\vert $. Thus, the usual adiabatic elimination technique does not
work well for this case. We note that the above Hamiltonian depicts a
typical spin-boson (continuous variable) system.

In the following parts of this paper, with the generalized BO approximation 
\cite{sun1}, and perturbation theory, we will prove that under the
conditions $\omega >>\omega _{a}$ and $\omega >>g$, the evolution of the two
qubit system is approximately governed by the effective Hamiltonian%
\begin{equation}
\hat{h}=\omega _{a}\left( \hat{\sigma}_{z}^{\left( 1\right) }+\hat{\sigma}%
_{z}^{\left( 2\right) }\right) -2\frac{g^{2}}{\omega }\hat{\sigma}%
_{x}^{\left( 1\right) }\hat{\sigma}_{x}^{\left( 2\right) }  \label{bbbbb}
\end{equation}%
and the decoherence effect caused by adiabatic entanglement \cite{sun2}
between the two qubits system and the data bus can be neglected.

Under the adiabatic condition that the external field does not excite the
transitions among the internal states of qubits, the BO approximaiton
determines a formally -factorized eigen-state of $H$ for the total composite
system. 
\begin{equation}
\left\langle \lambda_{1},\lambda_{2}\right. \left\vert \Psi \right\rangle
=\phi _{n\alpha }\left( \lambda_{1},\lambda_{2}\right) \left\vert
n[\lambda_{1},\lambda_{2}]\right\rangle  \label{ddd1}
\end{equation}%
where $\left\vert \lambda_{1},\lambda_{2}\right\rangle $ is the common
eigen-state $\left\vert \lambda_{1},\lambda_{2}\right\rangle $ of $\left\{ 
\hat{\sigma}_{x}^{\left( 1\right) },\hat{\sigma}_{x}^{\left( 2\right)
}\right\} $ with eigen-values $\left\{ \lambda_{1},\lambda_{2}=\pm 1\right\} 
$ and $\left\vert n[\lambda_{1},\lambda_{2}]\right\rangle $ the eigen-state
of the effective Hamiltonian 
\begin{equation}
H_{f}[\lambda_{1},\lambda_{2}]=\omega \hat{a}^{\dag }\hat{a}+g\left( \hat{a}%
^{\dag }+\hat{a}\right) \left( \lambda_{1}+\lambda_{2}\right)  \label{aaaaa}
\end{equation}%
of the external field for the given state $\left\vert \lambda _{1},\lambda
_{2}\right\rangle $ of the two qubit system with corresponding eigen-value 
\begin{eqnarray}
V_{n}\left( \lambda_{1},\lambda_{2}\right) &=&n\omega -\frac{g^{2}}{\omega }%
\left( \lambda_{1}+\lambda_{2}\right) ^{2}  \label{ccccc} \\
&=&n\omega -2\frac{g^{2}}{\omega }\lambda_{1}\lambda_{2}.  \nonumber
\end{eqnarray}%
Here, we have neglected a constant term. For different $n$, $V_{n}\left(
\lambda_{1},\lambda_{2}\right) $ contribute different effective potentials
for the slow part so that the wave function $\phi _{n\alpha }$ of the two
qubit system is determined by 
\begin{equation}
\hat{H}_{eff}(n)\left\vert \phi _{n\alpha }\right\rangle =E_{n\alpha
}\left\vert \phi _{n\alpha }\right\rangle  \label{eee1}
\end{equation}%
where the effective Hamiltonian is%
\begin{eqnarray}
\hat{H}_{eff}(n) &\equiv &\hat{H}_{eff}(n;\hat{\sigma}_{x}^{\left( 1\right)
},\hat{\sigma}_{x}^{\left( 2\right) })  \label{aa1} \\
&=&\omega _{a}\left( \hat{\sigma}_{z}^{\left( 1\right) }+\hat{\sigma}%
_{z}^{\left( 2\right) }\right) +V_{n}\left( \hat{\sigma}_{x}^{\left(
1\right) },\hat{\sigma}_{x}^{\left( 2\right) }\right)  \nonumber \\
&=&\hat{h}+n\omega  \nonumber
\end{eqnarray}%
where $\hat{h}$ defined in Eq. (\ref{bbbbb}) is the $n$-indepedent part of
the effective Hamiltonian.

So far, we have formally obtained a $X-X$-type of interbit coupling based on
the BO approximation, which can be conveniently used to formulate an
efficient scheme for quantum-computing. Recently, a similar interaction
generating the entanglement of two Josephson junction charge qubits was
proposed by Averin \cite{Averin}. In ref. \cite{Averin}, two Josephson
junction charge qubits are assumed to be coupled with another junction which
works as a faster data bus. With the BO approximation, the energy of the
lowest band of the latter junction can be considered as the effective
interaction between the two Josephson junction charge qubits.\ 

In the next section, we give a detailed derivation of the effective
Hamiltonian (\ref{aa1}) based on the perturbation theory. The decoherence of
the quantum state of the two qubit system induced by the adiabatic
entanglement with the external field is analyzed in section IV.

\section{Generalized Born-Oppenheimer Approximation for Discrete System}

\bigskip In the above section, we obtain the effective Hamiltonian (\ref%
{bbbbb}) for two qubit system through a intutitional analysis from the BO
approximation directly. However, the BO approximation is usually used to
seperate two continuous quantum systems (e.g. a melocular and a electron) or
a continuous system and a discrete one (e.g. the spatial motion and spin of
a neutral particle \cite{sun2}), rather than the two descrete subsystems. In
this section, however, we can generalize the BO approximaiton to deal with
the seperation of two discrete quantum systems, i.e. the two qubit system
and the external field. We now can give a rigorous proof to the intuituional
argument in the above section based on the perturbation theory. In this way,
we obtain the effective Hamiltonian (\ref{aa1}) naturally.

\subsection{Matrix Representation of Motion Equation}

In section II, we define $\left\vert n\left( \lambda _{1},\lambda
_{2}\right) \right\rangle $ as the eigen-state of the Hamiltonian $%
H_{f}[\lambda _{1},\lambda _{2}]$ of the external field and $\left\vert
\lambda _{1},\lambda _{2}\right\rangle $ the eigen-states of $\left\{ \hat{%
\sigma}_{x}^{\left( 1\right) },\hat{\sigma}_{x}^{\left( 2\right) }\right\} $%
. With the completeness relation \cite{sun1} 
\begin{equation}
\sum_{\lambda _{1},\lambda _{2}}\sum_{n}\left\vert n\left( \lambda
_{1},\lambda _{2}\right) \right\rangle \langle n\left( \lambda _{1},\lambda
_{2}\right) |\otimes \left\vert \lambda _{1},\lambda _{2}\right\rangle
\langle \lambda _{1},\lambda _{2}|=1,
\end{equation}%
the quantum state of the composite system can be expanded as 
\begin{equation}
\left\vert \Psi \right\rangle =\sum_{\left\{ n,\lambda _{1},\lambda
^{2}\right\} }C_{\lambda _{1},\lambda _{2}}\left( n\right) \left\vert n\left[
\lambda _{1},\lambda _{2}\right] \right)  \label{aa2}
\end{equation}%
where 
\[
\left\vert n\left[ \lambda _{1},\lambda _{2}\right] \right) \equiv
\left\vert n\left( \lambda _{1},\lambda _{2}\right) \right\rangle \left\vert
\lambda _{1},\lambda _{2}\right\rangle . 
\]%
Substituting Eq. (\ref{aa2}) into the Schr$\ddot{o}$dinger equation $%
H\left\vert \Psi \right\rangle =E\left\vert \Psi \right\rangle $, we obtain
the effective equations of the coefficients $C_{\lambda _{1},\lambda
_{2}}\left( n\right) $:%
\begin{eqnarray}
\sum_{\lambda _{1}^{\prime },\lambda _{2}^{\prime }}H_{\lambda _{1}^{\prime
},\lambda _{2}^{\prime }}^{\lambda _{1},\lambda _{2}}\left( n\right)
C_{\lambda _{1}^{\prime },\lambda _{2}^{\prime }}\left( n\right)
+\sum_{\lambda _{1}^{\prime },\lambda _{2}^{\prime }}F_{\lambda _{1}^{\prime
},\lambda _{2}^{\prime }}^{\lambda ^{1},\lambda _{2}}\left( n\right)
C_{\lambda _{1}^{\prime },\lambda _{2}^{\prime }}\left( n\right)  \label{o}
\\
+\sum_{\lambda _{1}^{\prime },\lambda _{2}^{\prime }}\sum_{m\neq
n}O_{\lambda _{1}^{\prime },\lambda _{2}^{\prime }}^{\lambda _{1},\lambda
_{2}}\left( n,m\right) C_{\lambda _{1}^{\prime },\lambda _{2}^{\prime
}}\left( m\right) =EC_{\lambda _{1},\lambda _{2}}\left( n\right)  \nonumber
\end{eqnarray}
where we have defined some complicated notations:%
\begin{equation}
H_{\lambda _{1}^{\prime },\lambda _{2}^{\prime }}^{\lambda _{1},\lambda
^{2}}\left( n\right) =\left\langle \lambda _{1},\lambda _{2}\right\vert \hat{%
H}_{eff}(n)\left\vert \lambda _{1}^{\prime },\lambda _{2}^{\prime
}\right\rangle  \label{ee1}
\end{equation}%
\begin{eqnarray}
F_{\lambda _{1}^{\prime },\lambda _{2}^{\prime }}^{\lambda _{1},\lambda
^{2}}\left( n\right) &=&\left\langle \lambda _{1},\lambda _{2}\right\vert
\omega _{a}\left( \hat{\sigma}_{z}^{\left( 1\right) }+\hat{\sigma}%
_{z}^{\left( 2\right) }\right) \left\vert \lambda _{1}^{\prime },\lambda
^{2\prime }\right\rangle  \label{dd1} \\
&&\times \left[ 1-\left\langle n(\lambda _{1},\lambda _{2})\right.
\left\vert n(\lambda _{1}^{\prime },\lambda _{2}^{\prime })\right\rangle %
\right]  \nonumber
\end{eqnarray}%
\begin{eqnarray}
O_{\lambda _{1}^{\prime },\lambda _{2}^{\prime }}^{\lambda _{1},\lambda
_{2}}\left( n,m\right) &=&\left\langle \lambda _{1},\lambda _{2}\right\vert
\omega _{a}\left( \hat{\sigma}_{z}^{\left( 1\right) }+\hat{\sigma}%
_{z}^{\left( 2\right) }\right) \left\vert \lambda _{1}^{\prime },\lambda
_{2}^{\prime }\right\rangle  \label{ff1} \\
&&\times \left\langle n(\lambda _{1},\lambda _{2})\right. \left\vert
m(\lambda _{1}^{\prime },\lambda _{2}^{\prime })\right\rangle .  \nonumber
\end{eqnarray}

Obviously, only when $F_{\lambda _{1}^{\prime },\lambda _{2}^{\prime
}}^{\lambda ^{1},\lambda _{2}}\left( n\right) $ and $O_{\lambda _{1}^{\prime
},\lambda ^{2\prime }}^{\lambda _{1},\lambda _{2}}\left( n,m\right) $ are
negligible, can we obtain the effective Hamiltonian $\hat{H}_{eff}(n)$. With
a straightforward calculation (see the Appendix A), $\left\langle n(\lambda
_{1},\lambda _{2})\right. \left\vert m(\lambda _{1}^{\prime },\lambda
_{2}^{\prime })\right\rangle $, $F_{\lambda _{1}^{\prime },\lambda
_{2}^{\prime }}^{\lambda ^{1},\lambda _{2}}\left( n\right) $ and $O_{\lambda
_{1}^{\prime },\lambda _{2}^{\prime }}^{\lambda _{1},\lambda _{2}}\left(
n,m\right) $ can be expressed as power series of the parameter $\frac{g}{%
\omega }$:%
\begin{eqnarray}
&&\left\langle m\left( \lambda _{1},\lambda _{2}\right) \right. \left\vert
n\left( \lambda _{1}^{\prime },\lambda _{2}^{\prime }\right) \right\rangle
\label{gg1} \\
&=&\exp \left[ -\frac{1}{2}\left\vert \frac{g}{\omega }\bar{\Sigma}_{\lambda
_{1}^{\prime },\lambda _{2}^{\prime }}^{\lambda _{1},\lambda
_{2}}\right\vert ^{2}\right] \times  \nonumber \\
&&\sum_{l=0}^{m}\frac{\left( -1\right) ^{n-m}\left( \frac{g}{\omega }\bar{%
\Sigma}_{\lambda _{1}^{\prime },\lambda _{2}^{\prime }}^{\lambda
_{1},\lambda ^{2}}\right) ^{2l+\left( n-m\right) }}{l!\left( l+n-m\right) !}%
\times  \nonumber \\
&&\sqrt{n\left( n-1\right) ...\left( m-l+1\right) }\sqrt{m\left( m-1\right)
..\left( m-l+1\right) },  \nonumber
\end{eqnarray}%
\begin{equation}
F_{\lambda _{1}^{\prime },\lambda _{2}^{\prime }}^{\lambda _{1},\lambda
^{2}}\left( n\right) =\omega _{a}\left[ (n+1)\frac{g^{2}}{\omega ^{2}}\bar{%
\Sigma}_{\lambda _{1}^{\prime },\lambda _{2}^{\prime }}^{\lambda
_{1},\lambda _{2}2}+...\right]  \label{hh1}
\end{equation}%
\begin{equation}
O_{\lambda _{1}^{\prime },\lambda _{2}^{\prime }}^{\lambda _{1},\lambda
^{2}}\left( n,m\right) =\omega _{a}\left[ -\sqrt{n}\frac{g}{\omega }\bar{%
\Sigma}_{\lambda _{1}^{\prime },\lambda _{2}^{\prime }}^{\lambda
_{1},\lambda ^{2}}\delta _{n,m\pm 1}+...\right]  \label{ii1}
\end{equation}%
where we have assumed $n-m\geq 0$ and $\bar{\Sigma}_{\lambda _{1}^{\prime
},\lambda _{2}^{\prime }}^{\lambda _{1},\lambda _{2}}=\lambda _{1}+\lambda
_{2}-\lambda _{1}^{\prime }-\lambda _{2}^{\prime }$. According to Eq. (\ref%
{hh1}), $F_{\lambda _{1}^{\prime },\lambda _{2}^{\prime }}^{\lambda
_{1},\lambda _{2}}\left( n\right) $ can be neglected if $n$ and $\frac{g}{%
\omega }$ are small enough. Considering the condition $\omega _{a}<<\omega $
mentioned in the above section, in the following discussion, we assume $%
n\lesssim 1$ , $g<<\omega $ and $\omega _{a}<<\omega $. For simplicity, we
also assume $g\symbol{126}\omega _{a}$ and $\frac{g}{\omega }\symbol{126}%
\frac{\omega _{a}}{\omega }$ is a small parameter.

In order to solve Eq. (\ref{o}) with the standard perturbation theory and
thus give a proof for the generalized BO approximation for the discrete
case, we rewrite Eq. (\ref{o}) in matrix-value form as%
\begin{equation}
\bar{H}C=EC  \label{jj1}
\end{equation}%
where the total Hamiltonian $\bar{H}$ and the eigen-vector $C$ are defined as%
\[
\bar{H}=\left[ 
\begin{array}{cc}
\bar{H}\left( 0\right) +\bar{F}\left( 0\right) , & \bar{O}\left( 0,1\right)
\\ 
\bar{O}\left( 1,0\right) , & \bar{H}\left( 1\right) +\bar{F}\left( 1\right)%
\end{array}%
\right] ,C=\left( 
\begin{array}{c}
C_{1} \\ 
C_{2}%
\end{array}%
\right) . 
\]%
Here, the vector $C$ contains two sub-vectors 
\[
C_{i}=\left( 
\begin{array}{cccc}
C_{11}\left( i\right) , & C_{1-1}\left( i\right) , & C_{-11}\left( i\right) ,
& C_{-1-1}\left( i\right)%
\end{array}%
\right) ^{T},i=0,1. 
\]%
Each of them is a $4$-dimension vector; $\bar{H}\left( i\right) $, $\bar{F}%
\left( i\right) $ and $\bar{O}\left( i,j\right) $ ($i,j=0,1$) are $4\times 4$
matrices whose elements are $H_{\lambda _{1}^{\prime },\lambda _{2}^{\prime
}}^{\lambda _{1},\lambda _{2}}\left( i\right) $, $F_{\lambda _{1}^{\prime
},\lambda _{2}^{\prime }}^{\lambda _{1},\lambda _{2}}\left( i\right) $ and $%
O_{\lambda _{1}^{\prime },\lambda _{2}^{\prime }}^{\lambda _{1},\lambda
_{2}}\left( i,j\right) $.

Making use of Eq. (\ref{hh1}) and Eq. (\ref{ii1}), we express the matrix $%
\bar{H}$ as a power series of $\frac{g}{\omega }$:%
\begin{equation}
\bar{H}\simeq \bar{H}^{\left( 0\right) }+\frac{g}{\omega }\bar{H}^{\left(
1\right) }+\left( \frac{g}{\omega }\right) ^{2}\left( \bar{V}^{\left(
2\right) }+\bar{O}^{\left( 2\right) }\right)  \label{kk1}
\end{equation}%
where 
\begin{equation}
\bar{H}^{\left( 0\right) }=\omega \left[ 
\begin{array}{cc}
0, & 0 \\ 
0, & I%
\end{array}%
\right] ,  \label{xx1}
\end{equation}%
\begin{equation}
\bar{H}^{\left( 1\right) }=\omega \left[ 
\begin{array}{cc}
\frac{\omega _{a}}{g}\left( \hat{\sigma}_{z}^{\left( 1\right) }+\hat{\sigma}%
_{z}^{\left( 2\right) }\right) , & 0 \\ 
0, & \frac{\omega _{a}}{g}\left( \hat{\sigma}_{z}^{\left( 1\right) }+\hat{%
\sigma}_{z}^{\left( 2\right) }\right)%
\end{array}%
\right] ,  \label{ww1}
\end{equation}%
\begin{equation}
\bar{V}^{\left( 2\right) }=\omega \left[ 
\begin{array}{cc}
\hat{\sigma}_{x}^{\left( 1\right) }\hat{\sigma}_{x}^{\left( 2\right) }, & 0
\\ 
0, & \hat{\sigma}_{x}^{\left( 1\right) }\hat{\sigma}_{x}^{\left( 2\right) }%
\end{array}%
\right] ,  \label{uu1}
\end{equation}%
and%
\begin{equation}
\bar{O}^{\left( 2\right) }=-\omega \left[ 
\begin{array}{cc}
0, & \frac{\omega _{a}}{g}\bar{\Sigma} \\ 
\frac{\omega _{a}}{g}\bar{\Sigma}, & 0%
\end{array}%
\right] .  \label{tt1}
\end{equation}%
Here, $I$ is the unit $4\times 4$ matrix and $\bar{\Sigma}$ a $4\times 4$
matrix whose elements are $\bar{\Sigma}_{\lambda _{1}^{\prime },\lambda
_{2}^{\prime }}^{\lambda _{1},\lambda _{2}}$. Only the terms of zeroth,
first and second order in $\frac{g}{\omega }$ are considered in Eq. (\ref%
{kk1}). The terms of higher order are neglected. According to Eq. (\ref{hh1}%
), $\bar{F}\left( i\right) $ ($i=0,1$) does not contain the terms of zeroth,
first and second order. Therefore, $\bar{F}\left( i\right) $ are neglected
in Eq. (\ref{kk1}).

\subsection{Perturbation Solution of Motion Equation}

To apply the perturbation theory, we split $\bar{H}$ into two parts:%
\begin{equation}
\bar{H}=\bar{H}_{0}+\bar{W}  \label{ll1}
\end{equation}%
where%
\begin{equation}
\bar{H}_{0}=\bar{H}^{\left( 0\right) }+\frac{g}{\omega }\bar{H}^{\left(
1\right) }  \label{mm1}
\end{equation}%
is treated as the unperturbated Hamiltonian and%
\begin{equation}
\bar{W}=\left( \frac{g}{\omega }\right) ^{2}\left( \bar{V}^{\left( 2\right)
}+\bar{O}^{\left( 2\right) }\right)  \label{nn1}
\end{equation}%
as the perturbation term.

It is pointed out that we can also choose $\bar{H}^{\left( 0\right) }$ as
the unperturbated Hamiltonian, $\frac{g}{\omega }\bar{H}^{\left( 1\right) }$
the first order perturbation term and $\left( \frac{g}{\omega }\right)
^{2}\left( \bar{V}^{\left( 2\right) }+\bar{O}^{\left( 2\right) }\right) $
the second order one. It is easy to prove that, if $\frac{g}{\omega }\bar{H}%
^{\left( 1\right) }$ is regarded as the first order perturbation
Hamiltonian, the eigenenergy and eigenstate to the first order given by the
perturbation theory is just the same as the ones we obtain by diagonalizing
the matrix $\bar{H}^{\left( 0\right) }+\frac{g}{\omega }\bar{H}^{\left(
1\right) }$ exactly. Therefore, our final result is not relevant to the
choice of unperturbated Hamiltonian and perturbation terms.

The eigen-energy and eigen-state of unperturbated Hamiltonian $\bar{H}_{0}$
in Eq. (\ref{mm1}) are easy to obtain. The eigen-energy can be expressed as:%
\begin{equation}
E_{ni}^{\left( 0\right) }=\varepsilon _{i}+n\omega  \label{aaaa1}
\end{equation}%
where $n=0,1$ and $i=0,+1,-1$. Here, we have defined 
\begin{equation}
\varepsilon _{0}=0\text{, \ \ }\varepsilon _{\pm 1}=\pm 2\omega _{a}\text{.}
\label{pp1}
\end{equation}%
The energy level structure of the compositive system is illustrated in Fig.
2. Since $\omega _{a}$ is much smaller than $\omega $, it follows from Eq. (%
\ref{aaaa1}) and Eq. (\ref{pp1}) that $E_{1i}^{\left( 0\right)
}-E_{0j}^{\left( 0\right) }\sim \omega $. Noting $\left\vert E_{ni}^{\left(
0\right) }-E_{nj}^{\left( 0\right) }\right\vert =\left\vert \varepsilon
_{i}-\varepsilon _{j}\right\vert \sim \omega _{a}$, we have $\left\vert
E_{ni}^{\left( 0\right) }-E_{nj}^{\left( 0\right) }\right\vert <<$ $%
E_{1i}^{\left( 0\right) }-E_{0j}^{\left( 0\right) }$. Therefore, the energy
spectrum of Hamiltonian $\bar{H}^{\left( 0\right) }$ possesses a qusi-band
structure: it consists of two qusi-bands corresponding to $n=0$ and $1$. 
\begin{figure}[h]
\begin{center}
\includegraphics[width=4.5cm,height=5cm]{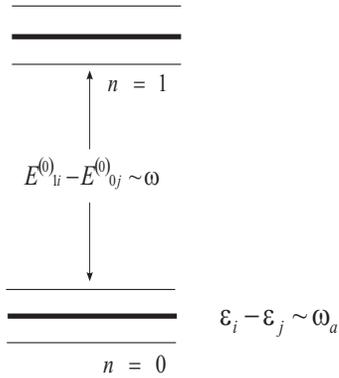} \vspace{0.3cm}
\end{center}
\caption{ The energy levels of $\bar{H}^{\left( 0\right) }$. Since $\protect%
\omega >>\protect\omega _{a},$ we have $E_{1i}^{\left( 0\right)
}-E_{0j}^{\left( 0\right) }>>\protect\varepsilon _{i}-\protect\varepsilon %
_{j}$. Therefore, there are two subbands with respect to $n=0$ and $n=1$.}
\end{figure}
The eigen-states corresponding to $E_{n\pm }$%
\begin{eqnarray}
\left\Vert E_{0\pm 1}^{\left( 0\right) }\right\rangle &=&\left( 
\begin{array}{cc}
C_{\pm }, & 0%
\end{array}%
\right) ^{T},  \nonumber \\
\left\Vert E_{0\pm 1}^{\left( 0\right) }\right\rangle &=&\left( 
\begin{array}{cc}
0, & C_{\pm }%
\end{array}%
\right) ^{T}  \label{qq1}
\end{eqnarray}%
can be written in terms of the unit vectors%
\begin{eqnarray*}
C_{+} &=&\left( 
\begin{array}{cccc}
1, & 0, & 0, & 0%
\end{array}%
\right) ^{T}, \\
C_{-} &=&\left( 
\begin{array}{cccc}
0, & 0, & 0, & 1%
\end{array}%
\right) ^{T}.
\end{eqnarray*}%
It is obvious from Eq. (\ref{xx1}) and Eq. (\ref{ww1}) that the
eigen-energies $E_{n0}^{\left( 0\right) }$ are two-fold degenerate, which
corresponds to the two degenerate eigen-vectors%
\begin{eqnarray}
\left\Vert E_{00;k}\right\rangle &=&\left( 
\begin{array}{cc}
C_{k}, & 0%
\end{array}%
\right) ^{T},  \nonumber \\
\left\Vert E_{10;k}\right\rangle &=&\left( 
\begin{array}{cc}
0, & C_{k}%
\end{array}%
\right) ^{T}.  \label{rr1}
\end{eqnarray}%
where $k=1,2$ and%
\begin{eqnarray*}
C_{1} &=&\left( 
\begin{array}{cccc}
0, & 1, & 0, & 0%
\end{array}%
\right) ^{T} \\
C_{2} &=&\left( 
\begin{array}{cccc}
0, & 0, & 1, & 0%
\end{array}%
\right) ^{T}.
\end{eqnarray*}

By making use of the perturbation theory, the eigen-energies and
eigen-states to the first order can be written as%
\[
E_{ni}=E_{ni}^{\left( 0\right) }+E_{ni}^{\left( 1\right) } 
\]%
and%
\begin{eqnarray*}
\left\Vert E_{n\pm 1}\right\rangle &=&\left\Vert E_{n\pm 1}^{\left( 0\right)
}\right\rangle +\left\Vert E_{n\pm 1}^{\left( 1\right) }\right\rangle \\
\left\Vert E_{n0;k}\right\rangle &=&\left\Vert E_{n0;k}^{\left( 0\right)
}\right\rangle +\left\Vert E_{n0;k}^{\left( 1\right) }\right\rangle .
\end{eqnarray*}%
According to the perturbation theory for degenerate levels, the zero-th
order eigen-states $\left\Vert E_{n0;k}^{\left( 0\right) }\right\rangle $
are linear combinations of $\left\Vert E_{n0;1}\right\rangle $ and $%
\left\Vert E_{n0;2}\right\rangle $. For any $k$ and $k^{\prime }$, $%
\left\langle E_{n0;1}\right\Vert \bar{O}^{\left( 2\right) }\left\Vert
E_{n0;2}\right\rangle =0$, $\left\Vert E_{n0;k}^{\left( 0\right)
}\right\rangle $ can be obtained by diagonalizing the $2\times 2$ matrices%
\begin{equation}
\left[ 
\begin{array}{cc}
\left\langle E_{n0;1}\right\Vert \bar{V}^{\left( 2\right) }\left\Vert
E_{n0;1}\right\rangle , & \left\langle E_{n0;1}\right\Vert \bar{V}^{\left(
2\right) }\left\Vert E_{n0;2}\right\rangle \\ 
\left\langle E_{n0;2}\right\Vert \bar{V}^{\left( 2\right) }\left\Vert
E_{n0;1}\right\rangle , & \left\langle E_{n0;2}\right\Vert \bar{V}^{\left(
2\right) }\left\Vert E_{n0;2}\right\rangle%
\end{array}%
\right] .  \label{ss1}
\end{equation}

The first order corrections to the eigen-energies $E_{ni}$ and eigen-states
can be calculated directly. Since we have $\left\langle E_{n\pm
1}\right\Vert \bar{O}^{\left( 2\right) }\left\Vert E_{n\pm 1}\right\rangle
=0 $ for every given $n$, it is obviously that the contributions from $\bar{O%
}^{\left( 2\right) }$ to the correction of the eigen-energies are zero. On 
 another hand, the first order eigen-vector $\left\Vert
E_{ni}^{\left( 1\right) }\right\rangle $ can be expressed as:%
\begin{eqnarray}
\left\Vert E_{ni}^{\left( 1\right) }\right\rangle &=&\left( \frac{g}{\omega }%
\right) ^{2}\frac{\left\langle E_{n-i}^{\left( 0\right) }\right\Vert \bar{V}%
^{\left( 2\right) }\left\Vert E_{ni}^{\left( 0\right) }\right\rangle }{%
\varepsilon _{i}-\varepsilon _{-i}}\left\Vert E_{n-i}^{\left( 1\right)
}\right\rangle  \label{oo1} \\
&&+\left( \frac{g}{\omega }\right) ^{2}\sum_{k=1,2}\frac{\left\langle
E_{n0;k}^{\left( 0\right) }\right\Vert \bar{V}^{\left( 2\right) }\left\Vert
E_{ni}^{\left( 0\right) }\right\rangle }{\varepsilon _{i}}\left\Vert
E_{n0;k}^{\left( 1\right) }\right\rangle  \nonumber \\
&&+\left( \frac{g}{\omega }\right) ^{2}\sum_{j=\pm 1}\frac{\left\langle
E_{\left\vert n-1\right\vert j}^{\left( 0\right) }\right\Vert \bar{O}%
^{\left( 2\right) }\left\Vert E_{ni}^{\left( 0\right) }\right\rangle }{%
\left( -1\right) ^{n+1}\omega +\varepsilon _{i}-\varepsilon _{j}}\left\Vert
E_{\left\vert n-1\right\vert j}^{\left( 1\right) }\right\rangle  \nonumber \\
&&+\left( \frac{g}{\omega }\right) ^{2}\sum_{k=1,2}\frac{\left\langle
E_{\left\vert n-1\right\vert 0;k}^{\left( 0\right) }\right\Vert \bar{O}%
^{\left( 2\right) }\left\Vert E_{ni}^{\left( 0\right) }\right\rangle }{%
\left( -1\right) ^{n+1}\omega +\varepsilon _{i}}\left\Vert E_{\left\vert
n-1\right\vert 0;k}^{\left( 1\right) }\right\rangle .  \nonumber
\end{eqnarray}%
The another eigenstate $\left\Vert E_{n0;k}^{\left( 1\right) }\right\rangle $
have the similar expression and thus for simplicity, we need not 
give its explicit expression here.

Substituting Eq. (\ref{pp1}-\ref{ss1}), into Eq. (\ref{oo1}), we can
calculate the contribution from $\bar{O}^{\left( 2\right) }$ and $\bar{V}%
^{\left( 2\right) }$ to the first order correction of
eigen-state. It is obvious that 
\begin{eqnarray}
&&\left( \frac{g}{\omega }\right) ^{2}\left\vert \frac{\left\langle
E_{n-i}^{\left( 0\right) }\right\Vert \bar{V}^{\left( 2\right) }\left\Vert
E_{ni}^{\left( 0\right) }\right\rangle }{\varepsilon _{i}-\varepsilon _{-i}}%
\right\vert  \nonumber \\
&\sim &\left( \frac{g}{\omega }\right) ^{2}\left\vert \frac{\left\langle
E_{n0;k}^{\left( 0\right) }\right\Vert \bar{V}^{\left( 2\right) }\left\Vert
E_{ni}^{\left( 0\right) }\right\rangle }{\varepsilon _{i}}\right\vert 
\nonumber \\
&\sim &\left( \frac{g}{\omega }\right) ^{2}\frac{\omega }{\omega _{a}}\sim 
\frac{g}{\omega }
\end{eqnarray}%
and%
\begin{eqnarray}
&&\left( \frac{g}{\omega }\right) ^{2}\left\vert \frac{\left\langle
E_{\left\vert n-1\right\vert j}^{\left( 0\right) }\right\Vert \bar{O}%
^{\left( 2\right) }\left\Vert E_{ni}^{\left( 0\right) }\right\rangle }{%
\left( -1\right) ^{n+1}\omega +\varepsilon _{i}-\varepsilon _{k}}\right\vert
\nonumber \\
&\sim &\left( \frac{g}{\omega }\right) ^{2}\left\vert \frac{\left\langle
E_{\left\vert n-1\right\vert 0,k}^{\left( 0\right) }\right\Vert \bar{O}%
^{\left( 2\right) }\left\Vert E_{ni}^{\left( 0\right) }\right\rangle }{%
\left( -1\right) ^{n+1}\omega +\varepsilon _{i}}\right\vert  \nonumber \\
&\sim &\left( \frac{g}{\omega }\right) ^{2}
\end{eqnarray}%
where we have used $\frac{g}{\omega _{a}}\symbol{126}1$. Therefore, it can
be seen from Eq. (\ref{oo1}) that the contribution from $\bar{O}^{\left(
2\right) }$ to $\left\Vert E_{ni}^{\left( 1\right) }\right\rangle $ and $%
\left\Vert E_{n0,k}^{\left( 1\right) }\right\rangle $ are much smaller than
the ones from $\bar{V}^{\left( 2\right) }$.

In summary, we have shown that one can reasonably neglect the first order
corrections from $\bar{O}^{\left( 2\right) }$ to both the eigen-values and
the eigen-states.

\subsection{The Generalized BO Approximation From Perturbation Theory}

The above result can be obtained with intutitve analysis. It can be seen
from Eq. (\ref{tt1}) and Eq. (\ref{uu1}) that the matrix $\bar{V}^{\left(
2\right) }$ leads to the transitions in the same energy band and $\bar{O}%
^{\left( 2\right) }$ leads to the interband transitions (see Fig. 3). 
\begin{figure}[h]
\begin{center}
\includegraphics[width=4.5cm,height=5cm]{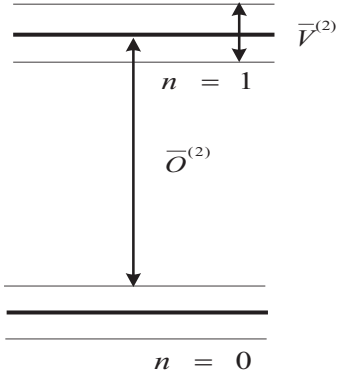} \vspace{0.3cm}
\end{center}
\caption{ The transitions induced by $\bar{W}$. We have shown that, $\left( 
\frac{g}{\protect\omega }\right) ^{2}\bar{O}^{\left( 2\right) }$ only
induces the transitions between subbands and $\left( \frac{g}{\protect\omega 
}\right) ^{2}\bar{V}^{\left( 2\right) }$ only induces the transitions in the
same subband. }
\end{figure}
As we have mentioned, the energy gap between the two bands has the same
order of $\omega $ and is much larger than the energy difference between the
levels in the same band. The latter have the same order of $\omega _{a}$.
Then the probabilities of the interband transitions are much smaller than
the ones of the transitions in the same energy band.

With the above discussions, the Hamiltonian matrix $\bar{H}$ can be written
as%
\begin{equation}
\bar{H}\approx \bar{H}^{\left( 0\right) }+\frac{g}{\omega }\bar{H}^{\left(
1\right) }+\left( \frac{g}{\omega }\right) ^{2}\bar{V}^{\left( 2\right) }
\label{vv1}
\end{equation}%
Which is just the effective Hamiltonian for the BO approximation. In fact,
substituting Eqs. (\ref{uu1}-\ref{xx1}) into Eq. (\ref{vv1}), we can write
the eigen equation (\ref{jj1}) as%
\[
\left[ 
\begin{array}{cc}
\begin{array}{c}
\omega _{a}\left( \hat{\sigma}_{z}^{\left( 1\right) }+\hat{\sigma}%
_{z}^{\left( 2\right) }\right) \\ 
+\frac{g^{2}}{\omega }\hat{\sigma}_{x}^{\left( 1\right) }\hat{\sigma}%
_{x}^{\left( 2\right) }%
\end{array}%
, & 0 \\ 
0, & 
\begin{array}{c}
\omega _{a}\left( \hat{\sigma}_{z}^{\left( 1\right) }+\hat{\sigma}%
_{z}^{\left( 2\right) }\right) \\ 
+\frac{g^{2}}{\omega }\hat{\sigma}_{x}^{\left( 1\right) }\hat{\sigma}%
_{x}^{\left( 2\right) }+\omega%
\end{array}%
\end{array}%
\right] C=EC. 
\]%
This equation equals to 
\begin{equation}
\sum_{\lambda _{1}^{\prime },\lambda _{2}^{\prime }}H_{\lambda _{1}^{\prime
},\lambda _{2}^{\prime }}^{\lambda _{1},\lambda _{2}}\left( n\right)
C_{\lambda _{1}^{\prime },\lambda _{2}^{\prime }}\left( n\right)
=EC_{\lambda _{1},\lambda _{2}}\left( n\right)  \label{yy1}
\end{equation}%
which is approximated form of Eq. (\ref{o}). It can be seen from Eq. (\ref%
{ee1}) that $H_{\lambda _{1}^{\prime },\lambda _{2}^{\prime }}^{\lambda
_{1},\lambda _{2}}\left( n\right) $ are the matrix elements of the effective
Hamiltonian $\hat{H}_{eff}(n)$ in Eq. (\ref{aa1}). It is pointed out that,
Eq. (\ref{yy1}) is equivalent with Eq. (\ref{ddd1}) and Eq. (\ref{eee1}) in
the above section.

To solve the time-dependent Schr$\ddot{o}$dinger equation which describe the
time evolution of the composite system, we can calculate the approximate
eigen-energies $E$ of the Hamiltonian of the composite system with Eq. (\ref%
{aa1}):%
\begin{equation}
E_{ni}=\epsilon _{i}+n\omega  \label{aaa1}
\end{equation}%
where $\epsilon _{i}$ is the $i$-th eigen-energy of the Hamiltonian $\hat{h}$
defined in Eq. (\ref{bbbbb}). Eq. (\ref{ee1}) is the time-independent Schr$%
\ddot{o}$dinger equation under the BO approximation. The time-dependent Schr$%
\ddot{o}$dinger equation can be written as:%
\begin{equation}
i\frac{d}{dt}C_{\lambda _{1},\lambda _{2}}\left( n\right) =\sum_{\lambda
_{1}^{\prime },\lambda _{2}^{\prime }}H_{\lambda _{1}^{\prime },\lambda
_{2}^{\prime }}^{\lambda _{1},\lambda _{2}}\left( n\right) C_{\lambda
_{1}^{\prime },\lambda _{2}^{\prime }}\left( n\right) .  \label{zz1}
\end{equation}

Making use of the above BO approximation for discrete system, we can
calculate the evolution of the composite system with Eq. (\ref{zz1}). If the
system is initially prepared in the state 
\[
\left\vert \Psi \left( 0\right) \right\rangle =\sum_{n,\left( \lambda
_{1},\lambda _{2}\right) }C_{\lambda _{1},\lambda _{2}}\left( n,0\right)
\left\vert n\left[ \lambda _{1},\lambda _{2}\right] \right) \text{,} 
\]%
then with Eq. (\ref{aaa1}), Eq. (\ref{aa1}) and Eq. (\ref{bbbbb}), the
quantum state at any instance can be expressed as: 
\begin{eqnarray}
\left\vert \Psi \left( t\right) \right\rangle &=&\sum_{n,\left( \lambda
_{1},\lambda _{2}\right) }C_{\lambda _{1},\lambda _{2}}\left( n,t\right)
e^{-in\omega t}\left\vert n\left[ \lambda _{1},\lambda _{2}\right] \right) 
\nonumber \\
&=&\sum_{n,\left( \lambda _{1},\lambda _{2}\right) }C_{\lambda _{1},\lambda
_{2}}\left( n,t\right) e^{-in\omega t}\left\vert \lambda _{1},\lambda
_{2}\right\rangle \otimes \left\vert n\left( \lambda _{1},\lambda
_{2}\right) \right\rangle  \label{ddddd}
\end{eqnarray}%
where 
\begin{eqnarray}
&&C_{\lambda _{1},\lambda _{2}}\left( n,t\right)  \nonumber \\
&=&\sum_{\lambda _{1}^{\prime },\lambda _{2}^{\prime }}\text{ }\left\langle
\lambda _{1},\lambda _{2}\right\vert e^{-i\hat{h}t}\left\vert \lambda
_{1}^{\prime },\lambda _{2}^{\prime }\right\rangle C_{\lambda _{1}^{\prime
},\lambda _{2}^{\prime }}\left( n,0\right) .  \label{x}
\end{eqnarray}%
Apparently, the evolution of the two qubit system is governed by the
Hamiltonian $\hat{h}$ in Eq. (\ref{bbbbb}) which contains the two qubit
interaction $-2\frac{g}{\omega ^{2}}\hat{\sigma}_{x}^{\left( 1\right) }\hat{%
\sigma}_{x}^{\left( 2\right) }$. However, it can be seen from Eq. (\ref%
{ddddd}) that because of the dependence of $\left\vert n\left( \lambda
_{1},\lambda _{2}\right) \right\rangle $ on $\left\{ \lambda _{1},\lambda
_{2}\right\} $, the instantaneous state $\left\vert \Psi \left( t\right)
\right\rangle $ of the composite system is usually an entangle state of the
two qubit system and the external field. Thus, if one trace over the
variable of the data bus, then {\bf the} the state of the two qubit system
is not a pure sate but a mixed one. This adiabatic decoherence effect will
be analyzed in next section.

\section{Decoherence Effect From the Adiabatic Entanglement}

In the above section, we have obtained an approximate Hamiltonian $\hat{H}%
_{eff}\left( n\right) =\hat{h}+n\omega $ for the composite system with two
qubits interacting with an external field. As we have mentioned, there is
quantum entanglement between the states of the external field and those of
the two qubit system because of the dependence of $\left\vert n\left(
\lambda _{1},\lambda _{2}\right) \right\rangle $ on $\left\{ \lambda
_{1},\lambda _{2}\right\} $. This is just the adiabatic quantum entanglement
which has been studied in detail for the development of BO approximation 
\cite{sun1}. It is easy to see that the adiabatic entanglement may cause the
adiabatic decoherence of the quantum state of the two qubit system. Only
when this decoherence effect can be neglected, can our effective Hamiltonian 
$\hat{h}$ be used to create a two bit gate. In this section, the adiabatic
decoherence effect in our problem will be analyzed in detail. It will be
shown that, in some practical cases, this kind of decoherence effect is
negligible and then the faster quantum entanglement can be created via the
BO approximation.

We assume the two qubit system and the external field are initially prepared
in a factorizable state: 
\[
\left\vert \Psi \left( 0\right) \right\rangle =\left\vert \psi \right\rangle
\otimes \left\vert \phi \right\rangle 
\]%
where $\left\vert \psi \right\rangle $ is the initial state of the two
qubits and $\left\vert \phi \right\rangle $ the state of the external field. 
$\left\vert \Psi \left( 0\right) \right\rangle $ can be expanded with
respect to the basis $\left\vert n\left[ \lambda _{1},\lambda _{2}\right]
\right) $ as 
\[
\left\vert \Psi \left( 0\right) \right\rangle =\sum_{n,\left( \lambda
_{1},\lambda _{2}\right) }C_{\lambda _{1},\lambda _{2}}\left( n,0\right)
\otimes \left\vert n\left[ \lambda _{1},\lambda _{2}\right] \right) 
\]%
or equivalently in the $\left\vert \lambda _{1},\lambda _{2}\right\rangle $
picutre 
\[
\left\langle \lambda _{1},\lambda _{2}\right. \left\vert \Psi \left(
0\right) \right\rangle =\sum_{n}C_{\lambda _{1},\lambda _{2}}\left(
n,0\right) \otimes \left\vert n\left[ \lambda _{1},\lambda _{2}\right]
\right\rangle 
\]%
where 
\begin{equation}
C_{\lambda _{1},\lambda _{2}}\left( n,0\right) =\langle \lambda _{1},\lambda
_{2}\left\vert \psi \right\rangle \cdot \langle n(\lambda _{1},\lambda
_{2})\left\vert \phi \right\rangle .  \label{nn}
\end{equation}%
is the projection of inital state onto the completeness basis $\left\vert n%
\left[ \lambda _{1},\lambda _{2}\right] \right\rangle $. With the BO
approximation generalized in the above section, the evolution of the
composite system is depicted by Eq. (\ref{ddddd}).

Substituting Eq. (\ref{nn}) into Eq. (\ref{ddddd}), we have the wave
function in the{\bf \ }$\left( \lambda _{1},\lambda _{2}\right) $ picture at
time $t$. 
\begin{eqnarray}
&&\langle \lambda _{1},\lambda _{2}\left\vert \Psi \left( t\right)
\right\rangle =  \nonumber \\
&&\sum_{\lambda _{1}^{\prime },\lambda _{2}^{\prime }}\text{ }\left\langle
\lambda _{1},\lambda _{2}\right\vert e^{-i\hat{h}t}\left\vert \lambda
_{1}^{\prime },\lambda _{2}^{\prime }\right\rangle \langle \lambda
_{1}^{\prime },\lambda _{2}^{\prime }\left\vert \psi \right\rangle \otimes
\left\vert \phi \left( t\right) \right\rangle
\end{eqnarray}%
where 
\begin{eqnarray*}
\left\vert \phi \left( t\right) \right\rangle &=&\left\vert \phi \left(
t,\lambda _{1},\lambda _{2},\lambda _{1}^{\prime },\lambda _{2}^{\prime
}\right) \right\rangle \\
&=&\sum_{n}\text{ }e^{-in\omega t}\langle n(\lambda _{1}^{\prime },\lambda
_{2}^{\prime })\left\vert \phi \right\rangle \left\vert n(\lambda
^{1},\lambda _{2})\right\rangle
\end{eqnarray*}%
depends on both coordinate sets ($\lambda _{1},\lambda _{2}$) and ($\lambda
_{1}^{\prime },\lambda _{2}^{\prime }$). It should be noted that the factor $%
\langle n(\lambda _{1}^{\prime },\lambda _{2}^{\prime })\left\vert \phi
\right\rangle \left\vert n(\lambda _{1},\lambda _{2})\right\rangle $ has
different pairs of indices ($\lambda _{1},\lambda _{2}$) and ($\lambda
_{1}^{\prime },\lambda _{2}^{\prime }$). If $\left\vert \phi \left( t\right)
\right\rangle =\left\vert \phi \left( t,\lambda _{1},\lambda _{2},\lambda
_{1}^{\prime },\lambda _{2}^{\prime }\right) \right\rangle $ were
independent of $\lambda _{1},\lambda _{2}$ and $\lambda _{1}^{\prime }$,$%
\lambda _{2}^{\prime }$, then we would have a factorized evolution described
by $\left\vert \Psi \left( t\right) \right\rangle =e^{-i\hat{h}t}\left\vert
\psi \right\rangle \otimes \left\vert \phi \left( t\right) \right\rangle $.
In this case, the two qubit system were just in the pure state controlled by
the effective Hamiltonian $\hat{h}$ and the decoherence phenomenon does not
occur.

In the{\bf \ }usual case, $\left\vert \phi \left( t\right) \right\rangle
=\left\vert \phi \left( t,\lambda _{1},\lambda _{2},\lambda _{1}^{\prime
},\lambda _{2}^{\prime }\right) \right\rangle $ does depend on $\lambda
_{1},\lambda _{2},\lambda _{1}^{\prime }$ and $\lambda _{2}^{\prime }$. Then
the decoherence effect must result from the dependence $\left\vert \phi
\left( t\right) \right\rangle $ to different sets of ($\lambda _{1},\lambda
_{2},\lambda _{1}^{\prime },\lambda _{2}^{\prime }$). In fact, we can
calculate the elements of the reduced density matrix of the two qubit \
system

\begin{eqnarray}
&&\rho _{\lambda _{1},\lambda _{2},\lambda _{1}^{\prime },\lambda
_{2}^{\prime }}\left( t\right) =  \nonumber \\
&&\sum_{\mu _{1},\mu _{2}}\sum_{\mu _{1}^{\prime },\mu _{2}^{\prime
}}K_{\lambda _{1},\lambda _{2};\mu _{1},\mu _{2}}[\left\vert \psi
\right\rangle ,t]K_{\lambda _{1}^{\prime },\lambda _{2}^{\prime };\mu
_{1}^{\prime },\mu _{2}^{\prime }}^{\ast }[\left\vert \psi \right\rangle ,t]
\nonumber \\
&&\times \langle \phi \left( t;\lambda _{1}^{\prime },\lambda _{2}^{\prime
};\mu _{1}^{\prime },\mu _{2}^{\prime }\right) \left\vert \phi \left(
t,\lambda _{1},\lambda _{2};\mu _{1},\mu _{2}\right) \right\rangle
\label{cccc1}
\end{eqnarray}%
where the propagation functionals 
\begin{equation}
K_{\lambda _{1},\lambda _{2};\mu _{1},\mu _{2}}[\left\vert \psi
\right\rangle ,t]=\left\langle \lambda _{1},\lambda _{2}\right\vert e^{-i%
\hat{h}t}\left\vert \mu _{1},\mu _{2}\right\rangle \langle \mu _{1},\mu
_{2}\left\vert \psi \right\rangle .
\end{equation}%
depend on the initial state $\left\vert \psi \right\rangle $ of the external
field. On the other hand, without decoherence, the reduced density matrix of
the qubits is $\rho ^{\prime }=e^{-i\hat{h}t}\left\vert \psi \right\rangle
\left\langle \psi \right\vert e^{i\hat{h}t}$ with matrix elements 
\begin{eqnarray}
&&\rho _{\lambda _{1},\lambda _{2},\lambda _{1}^{\prime },\lambda
_{2}^{\prime }}^{\prime }\left( t\right)  \label{bbbb1} \\
&=&\sum_{\mu _{1},\mu _{2}}\sum_{\mu _{1}^{\prime },\mu _{2}^{\prime
}}K_{\lambda _{1},\lambda _{2};\mu _{1},\mu _{2}}[\left\vert \psi
\right\rangle ,t]K_{\lambda _{1}^{\prime },\lambda _{2}^{\prime };\mu
_{1}^{\prime },\mu _{2}^{\prime }}^{\ast }[\left\vert \psi \right\rangle ,t]
\nonumber
\end{eqnarray}%
Comparing Eq. (\ref{bbbb1}) with Eq. (\ref{cccc1}), we measure the extent of
decoherence by the so called decoherence factor \cite{sun3}. 
\begin{equation}
F\equiv \langle \phi \left( t;\lambda _{1}^{\prime },\lambda _{2}^{\prime
};\mu _{1}^{\prime },\mu _{2}^{\prime }\right) \left\vert \phi \left(
t,\lambda _{1},\lambda _{2};\mu _{1},\mu _{2}\right) \right\rangle
\end{equation}%
As pointed in ref. \cite{sun3}, if $F=1$, there is no decoherence. It should
be pointed out that, not only the norm of $F$ but also its phase are
crucially important to the description of quantum decoherence process. For
example, if $F=\exp \left[ i\Theta \right] $ and $\Theta $ is a random phase
with respect to the parameters $\lambda _{1},\lambda _{2},\lambda
_{1}^{\prime },\lambda _{2}^{\prime };\mu _{1}^{\prime },\mu _{2}^{\prime
},;\mu ^{1},\mu ^{2}$, even if $\left\vert F\right\vert =1$, $\rho _{\lambda
_{1},\lambda _{2},\lambda _{1}^{\prime },\lambda _{2}^{\prime }}^{\prime
}\left( t\right) $ still have significant difference from the pure $\rho
_{\lambda _{1},\lambda _{2},\lambda _{1}^{\prime },\lambda _{2}^{\prime
}}\left( t\right) $ .

We assume the initial state of the external field is in a coherence state, $%
\left\vert \phi \right\rangle =\left\vert \alpha \right\rangle $ (noting the
condition $n\lesssim 1$ in the above section, we assume $\left\vert \alpha
\right\vert <0.5$ here). By a straightforward calculation, we obtain%
\begin{eqnarray}
&&\left\vert \phi \left( t,\lambda _{1},\lambda _{2},\lambda _{1}^{\prime
},\lambda _{2}^{\prime }\right) \right\rangle  \nonumber \\
&=&e^{i\Phi \left( t,\lambda _{1},\lambda _{2},\lambda _{1}^{\prime
},\lambda _{2}^{\prime }\right) }\times  \nonumber \\
&&\left\vert \left[ \alpha +\frac{g}{\omega }\left( \lambda _{1}^{\prime
}+\lambda _{2}^{\prime }\right) \right] e^{-i\omega t}-\frac{g}{\omega }%
\left( \lambda _{1}+\lambda _{2}\right) \right\rangle
\end{eqnarray}%
where the time-varing phase%
\begin{eqnarray}
\Phi &=&\left[ \frac{g}{\omega }\left( \lambda _{1}^{\prime }+\lambda
_{2}^{\prime }\right) \right] \text{Im}\alpha ^{\ast }  \nonumber \\
&&-\left[ \frac{g}{\omega }\left( \lambda _{1}+\lambda _{2}\right) \right] 
\text{Im}\left( \alpha ^{\ast }e^{i\omega t}\right)  \nonumber \\
&&-\left[ \left( \frac{g}{\omega }\right) ^{2}\left( \lambda _{1}^{\prime
}+\lambda _{2}^{\prime }\right) \left( \lambda _{1}+\lambda _{2}\right) %
\right] \text{sin}\omega t
\end{eqnarray}%
depends on the parameters $\lambda _{1}^{\prime },\lambda _{2}^{\prime },$ $%
\lambda _{1}$ and $\lambda _{2}$ and $\left\vert \left[ \alpha +\frac{g}{%
\omega }\left( \lambda _{1}^{\prime }+\lambda _{2}^{\prime }\right) \right]
e^{-i\omega t}-\frac{g}{\omega }\left( \lambda _{1}+\lambda _{2}\right)
\right\rangle $ is also a coherent state.

Considering the inner product of coherent state satisfies%
\begin{equation}
\left\langle \alpha \right. \left\vert \alpha +p\right\rangle =\exp \left[
-\left\vert p\right\vert ^{2}\right] \exp \left[ i\text{Im}\left( p\alpha
^{\ast }\right) \right] ,  \nonumber
\end{equation}%
we can express the decoherence factor 
\begin{equation}
F=\exp \left[ i\Omega \right] \cdot \exp \left[ -\left\vert \Upsilon
\right\vert ^{2}\right]  \nonumber
\end{equation}%
in terms of the phase factor $\Omega $ and its norm $\exp \left[ -\left\vert
\Upsilon \right\vert ^{2}\right] .$ Here, the explicit expression of $\Omega 
$ and $\Upsilon $ can be obtained: 
\begin{eqnarray}
\Omega &=&2\left[ \frac{g}{\omega }\left( \mu _{1}+\mu _{2}-\mu _{1}^{\prime
}-\mu _{2}^{\prime }\right) \right] \text{Im}\alpha ^{\ast }  \nonumber \\
&&+2\left[ \frac{g}{\omega }\left( \lambda _{1}^{\prime }+\lambda
_{2}^{\prime }-\lambda _{1}-\lambda _{2}\right) \right] \text{Im}\left(
\alpha ^{\ast }e^{i\omega t}\right)  \nonumber \\
&&+\left( \frac{g}{\omega }\right) ^{2}\left( \mu _{1}+\mu _{2}-\mu
_{1}^{\prime }-\mu _{2}^{\prime }\right)  \nonumber \\
&&\times \left( \lambda _{1}^{\prime }+\lambda _{2}^{\prime }+\lambda
_{1}+\lambda _{2}\right) \text{sin}\omega t  \label{zz}
\end{eqnarray}%
and 
\begin{eqnarray}
\Upsilon &=&\left[ \frac{g}{\omega }\left( \mu _{1}+\mu _{2}-\mu
_{1}^{\prime }-\mu _{2}^{\prime }\right) \right] e^{-i\omega t}  \nonumber \\
&&-\frac{g}{\omega }\left( \lambda _{1}+\lambda _{2}-\lambda _{1}^{\prime
}-\lambda _{2}^{\prime }\right) .
\end{eqnarray}

However, in the case with 
\begin{equation}
\left\vert \Omega \right\vert <<1,\left\vert \Upsilon \right\vert ^{2}<<1
\label{auu}
\end{equation}%
we have $\langle \phi \left( t;\lambda _{1}^{\prime },\lambda _{2}^{\prime
};\mu _{1}^{\prime },\mu _{2}^{\prime }\right) \left\vert \phi \left(
t,\lambda _{1},\lambda _{2};\mu _{1},\mu _{2}\right) \right\rangle \simeq 1$
and the decoherence effect can be neglected.

It is observed from the explicit expression for the decoherence factor that $%
\Omega $ and $\Upsilon $ are all periodic functions of $t$ with frequency $%
\omega $. The amplitudes of $\Omega $ and $\Upsilon $ do not vary with time
(see Fig. 4). Therefore, in case the amplitudes of $\Omega $ and $\Upsilon $
are much smaller than $1$, the inequalities in Eq. (\ref{auu}) can be
satisfied at any instance. 
\begin{figure}[h]
\begin{center}
\includegraphics[width=7.5cm,height=5cm]{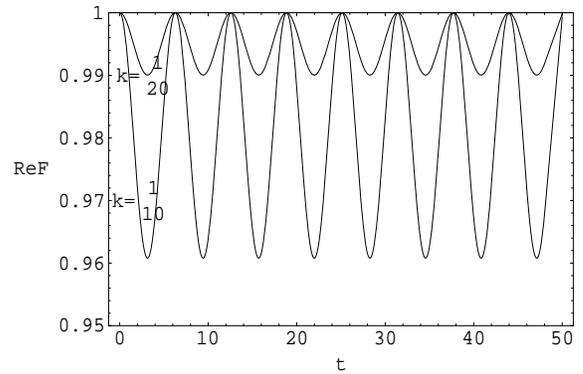} \vspace{0.3cm}
\end{center}
\caption{ The real part of the decoherence factor $\text{Re}F$ as a function
of $t$. The unit of $t$ is $\protect\omega ^{-1}$. We assume $\protect\alpha %
=0$ and $\protect\mu ^{1}+\protect\mu ^{2}-\protect\mu _{1}^{\prime }-%
\protect\mu ^{2^{\prime }}=\protect\lambda _{1}+\protect\lambda _{2}-\protect%
\lambda ^{1^{\prime }}-\protect\lambda ^{2^{\prime }}=1.$ The ratio $\frac{g%
}{\protect\omega }$ is denoted as $k$. It can be seen from the figure that
the decoherence factor is a periodic function of time.}
\end{figure}

Because every $\lambda_{1}$ or $\lambda_{2}$ may be $\pm 1$, the sufficient
condition of $\left\vert \Upsilon \right\vert ^{2}<<1$ is just $\left\vert 8%
\frac{g}{\omega }\right\vert ^{2}<<1$ and if $\frac{g}{\omega }$ is small
enough so that this condition can be satisfied, then the last term of the
expression of $\Omega $ can be neglected and the sufficient condition of $%
\left\vert \Omega \right\vert <<1$ becomes 
\begin{equation}
16\frac{g}{\omega }\left\vert \alpha \right\vert <<1.  \label{ww}
\end{equation}%
This is a restriction for the strength of the external field. Therefore, we
know that if the ratio $\frac{g}{\omega }$ is small enough (for example, $%
\frac{g}{\omega }\sim 0.01$) and the external field is weak enough (e.g. $%
\left\vert \alpha \right\vert <<1$), the atomic state decoherence caused by
the external field can be omitted and the two qubit system can evolve under
the control of the effective Hamiltonian $\hat{h}$.

\section{Conclusion with Remarks}

In this paper, we have proposed a new protocol for the creation of quantum entanglement between two qubits interacting with a high frequency data bus. Unlike the previous schemes,
our protocal does not require the rotation wave approximation for the
interaction Hamiltonian. We can first assume the external field is prepared
in the coherence state $\left\vert \alpha \right\rangle $. If the frequency
of the external field is large enough and the strength of it is weak
enough so that the following conditions%
\begin{eqnarray}
\omega _{a} &\sim &g<<\omega \text{, \ \ }64\frac{g^{2}}{\omega ^{2}}<<1, 
\nonumber \\
\left\vert \alpha \right\vert &<&0.5\text{, \ \ }\frac{16g}{\omega }%
\left\vert \alpha \right\vert <<1  \label{ab}
\end{eqnarray}%
can be satisfied, an effective coupling between the two qubit system{\bf \ }%
can be obtained under the BO approximation and the decoherence caused by the
cavity field can be neglected.

The explicit expression of the Hamiltonian of the composite system is given
in Eq. (\ref{a}). If the conditions by Eq. (\ref{ab}) are satisfied, the
evolution of the two qubit system is governed by the effective Hamiltonian%
\begin{equation}
\hat{h}=\omega _{a}\left( \hat{\sigma}_{z}^{\left( 1\right) }+\hat{\sigma}%
_{z}^{\left( 2\right) }\right) -2\frac{g^{2}}{\omega }\hat{\sigma}%
_{x}^{\left( 1\right) }\hat{\sigma}_{x}^{\left( 2\right) }\text{.} \label{ba}
\end{equation}

In our previous calculation, we have assumed $\omega _{a}\sim g<<\omega $.
This is the sufficient condition to obtain the effective Hamiltonian (\ref%
{ba}). It is easy to prove that if $\omega _{a}\ $is smaller than $g$ and $%
g<<\omega $, the effective Hamiltonian (\ref{ba}) can be also obtained with
the deduction in the above sections. To realize a logic gate with high
efficiency, we suppose our proposal work under the condition $\omega _{a}<<g$
and $g<<\omega $ so that the interaction strength $2\frac{g^{2}}{\omega }$
is comparable with the strength $\omega _{a}$ of the free Hamiltonian.

As we have shown, the decoherence effect is determined by the factor $\Omega 
$ and $\left\vert \Upsilon \right\vert ^{2}$ which are periodic functions of
time. Then if the conditions in Eq. (\ref{ab}) are satisfied, the
decoherence effect can be neglected no matter how long the operation takes.

The scheme proposed in this paper can be generalized to the system that $N$
qubits coupled to a data bus. With the same discussions as above, it is
apparently that in the large detuning and weak coupling case where $\omega
_{a}<<\omega $, $g<<\omega $, and the number of the qubits is not very large
so that the condition $2N\omega _{a}<<\omega $ is satisfied, an effective
coupling between the qubits 
\[
V\sim -\frac{g^{2}}{\omega }\left( \hat{\sigma}_{x}^{\left( 1\right) }+\hat{%
\sigma}_{x}^{\left( 2\right) }+...+\hat{\sigma}_{x}^{\left( N\right)
}\right) ^{2}=-\frac{g^{2}}{\omega }J_{x}^{2} 
\]%
can be obtained under the BO approximation. It is pointed out that with the "%
$J_{x}^{2}$ interaction" between $N$ qubits, the $N$-qubit GHZ state can be
created easily (\cite{Jx}). Then a physical qubit of the form $\alpha
\left\vert 0\right\rangle +\beta \left\vert 1\right\rangle $ can be encoded
into a logical qubit of the form $\alpha \left\vert 000...\right\rangle
+\beta \left\vert 111...\right\rangle $ (\cite{zeng}). It is also proved
that with the interaction of this type, Shor code for error correction can
be easily realized.

{\bf Acknowledgement:} {\it \ This work is supported by the NSFC (grant
No.90203018) and the knowledge Innovation Program (KIP) of the Chinese
Academy of Sciences and the National Fundamental Research Program of China
with No 001GB309310. We also sincerely thank L. You for the helpful
discussions with him.}

\appendix

\section{The Calculation of The Overlap Factor $\left\langle m\left( \protect%
\lambda _{1},\protect\lambda _{2}\right) \right. \left\vert n\left( \protect%
\lambda _{1}^{\prime },\protect\lambda _{2}^{\prime }\right) \right\rangle $}

In this appendix, we calculate the explicit expression of $\left\langle
m\left( \lambda _{1},\lambda _{2}\right) \right. \left\vert n\left( \lambda
_{1}^{\prime },\lambda _{2}^{\prime }\right) \right\rangle $. It can be seen
from Eq. (\ref{aaaaa}) that the Hamiltonian $H_{f}\left( \lambda
_{1},\lambda _{2}\right) $ is just the Hamiltonian of a displaced harmonic
oscillator and can be expressed as%
\begin{equation}
H_{f}\left( \lambda _{1},\lambda _{2}\right) \simeq \omega \hat{A}_{\lambda
_{1},\lambda _{2}}^{\dagger }\hat{A}_{\lambda _{1},\lambda _{2}}-2\frac{g^{2}%
}{\omega }\lambda _{1}\lambda _{2}  \label{bbb1}
\end{equation}%
where the constant term independent on $\lambda _{1}$,$\lambda _{2}$ is
omitted. Here,%
\[
\hat{A}_{\lambda _{1},\lambda _{2}}=\hat{a}+\frac{g}{\omega }\left( \lambda
_{1}+\lambda _{2}\right) 
\]%
is a displaced boson operator that satisfies $\left[ \hat{A}_{\lambda
_{1},\lambda _{2}},\hat{A}_{\lambda _{1},\lambda _{2}}^{\dag }\right] =1.$
With Eq. (\ref{bbb1}), the eigen-state of $H_{f}\left( \lambda _{1},\lambda
_{2}\right) $ can be written as:%
\begin{eqnarray}
\left\vert n(\lambda _{1},\lambda _{2})\right\rangle  &=&\frac{1}{\sqrt{n!}}%
\hat{A}_{\lambda _{1},\lambda _{2}}^{\dagger n}\left\vert 0\left( \lambda
_{1},\lambda _{2}\right) \right\rangle   \nonumber \\
&=&\hat{D}\left[ -\frac{g}{\omega }\left( \lambda _{1}+\lambda _{2}\right) %
\right] \left\vert n\right\rangle   \label{bbb2}
\end{eqnarray}%
in terms of the displacement operator $\hat{D}\left[ -\frac{g}{\omega }%
\left( \lambda _{1}+\lambda _{2}\right) \right] $ and the natural Fock state 
$\left\vert n\right\rangle =$ $\frac{1}{\sqrt{n!}}\hat{a}^{\dagger
n}\left\vert 0\right\rangle $. Here, $\left\vert 0\right\rangle $ is the
realistic vacuum state satisfies $\hat{a}\left\vert 0\right\rangle =0$ and $%
\left\vert 0\left( \lambda _{1},\lambda _{2}\right) \right\rangle $ the
displaced vacuum state satisfies $\hat{A}_{\lambda _{1},\lambda
_{2}}\left\vert 0\left( \lambda _{1},\lambda _{2}\right) \right\rangle =0$.
It is easy to proof that $\left\vert 0\left( \lambda _{1},\lambda
_{2}\right) \right\rangle $ is{\bf \ }actually a coherence state which can
be defined as $\left\vert 0\left( \lambda _{1},\lambda _{2}\right)
\right\rangle =\hat{D}\left[ -\frac{g}{\omega }\left( \lambda _{1}+\lambda
_{2}\right) \right] \left\vert 0\right\rangle $.

With Eq. (\ref{bbb2}), the explicit value of $\left\langle m\left( \lambda
_{1},\lambda _{2}\right) \right. \left\vert n\left( \lambda _{1}^{\prime
},\lambda _{2}^{\prime }\right) \right\rangle $ can be obtained easily:%
\begin{eqnarray*}
&&\left\langle m\left( \lambda _{1},\lambda _{2}\right) \right. \left\vert
n\left( \lambda _{1}^{\prime },\lambda _{2}^{\prime }\right) \right\rangle
=\left\langle m\right\vert D\left[ \frac{g}{\omega }\bar{\Sigma}_{\lambda
_{1}^{\prime },\lambda _{2}^{\prime }}^{\lambda _{1},\lambda _{2}}\right]
\left\vert n\right\rangle \\
&=&e^{-\frac{1}{2}\left\vert \frac{g}{\omega }\bar{\Sigma}_{\lambda
_{1}^{\prime },\lambda _{2}^{\prime }}^{\lambda _{1},\lambda
_{2}}\right\vert ^{2}}\left\langle m\right\vert e^{\frac{g}{\omega }\bar{%
\Sigma}_{\lambda _{1}^{\prime },\lambda _{2}^{\prime }}^{\lambda
_{1},\lambda _{2}}a^{+}}e^{-\frac{g}{\omega }\bar{\Sigma}_{\lambda
_{1}^{\prime },\lambda _{2}^{\prime }}^{\lambda _{1},\lambda
_{2}}a}\left\vert n\right\rangle \\
&=&e^{-\frac{1}{2}\left\vert \frac{g}{\omega }\bar{\Sigma}_{\lambda
_{1}^{\prime },\lambda _{2}^{\prime }}^{\lambda _{1},\lambda
_{2}}\right\vert ^{2}}\sum_{l=0}^{m}\frac{\left( -1\right) ^{n-m}\left( 
\frac{g}{\omega }\bar{\Sigma}_{\lambda _{1}^{\prime },\lambda _{2}^{\prime
}}^{\lambda _{1},\lambda _{2}}\right) ^{2l+\left( n-m\right) }}{l!\left(
l+n-m\right) !} \\
&&\sqrt{n\left( n-1\right) ...\left( m-l+1\right) }\sqrt{m\left( m-1\right)
..\left( m-l+1\right) }.
\end{eqnarray*}%
This is just Eq. (\ref{gg1}) in section III. Substituting Eq. (\ref{gg1})
into Eq. (\ref{dd1}) and Eq. (\ref{ff1}), we can get the expression of $%
F_{\lambda _{1}^{\prime },\lambda _{2}^{\prime }}^{\lambda _{1},\lambda
_{2}}\left( n\right) $ and $O_{\lambda _{1}^{\prime },\lambda _{2}^{\prime
}}^{\lambda _{1},\lambda _{2}}\left( n\right) $ i.e. Eq. (\ref{hh1}) and Eq.
(\ref{ii1}).


\begin{references}
\bibitem[a]{email} Electronic address: suncp@itp.ac.cn

\bibitem[b]{www} Internet www site: http:// www.itp.ac.cn/~suncp



\bibitem{Gate2} A. Barenco, C. H. Bennett, R. Cleve, D. P. DiVincenzo, N.
Margolus, P. Shor, T. Sleator, J. A. Smolin, and H. Weinfurter, Phys. Rev. A 
{\bf 52}, 3457 (1995).

\bibitem{Gate} A. Barenco, D. Deustch, and A. Ekert, Phys. Rev. Lett. {\bf 74%
}, 4083 (1995).

\bibitem{effort1} Q. A. Turchette, C. J. Hood, W. Lange, H. Mabuchi, and H.
J. Kimble, Phys. Rev. Lett. {\bf 75}, 4710 (1995); A. rauschenbeutel, G.
Nogues, S. Osnaghi, P. Bertet, M. Brune, J. M. Raimond, and S. Haroche,
Phys. Rev. Lett. {\bf 83}, 5166 (1999); S. Osnaghi, P. Bertet, A. Auffeves,
P. Maioli, M. Brune, J. M. Raimond, and S. Haroche, Phys. Rev. Lett. {\bf 87}%
, 037902 (2001).

\bibitem{effort2} C. Monroe, D. M. Meekhof, B. E. King, W. M. Itano, and D.
J. Wineland, Phys. Rev. Lett. {\bf 75}, 4714 (1995).

\bibitem{effort3} N. A. Gershenfeld and I. L. Chuand, Science {\bf 275}, 350
(1997); D. G. Cory, A. F. Fahmy, and T. F. Havel, Proc. Natl. Acad. Sci.
U.S.A. {\bf 94}, 1634 (1997); J. A. Jones, M. Mosca, and R. H. Hansen,
Nature (London) {\bf 393}, 344 (1998).

\bibitem{effort4} Yu. A. Pashkin, T. Yamamoto, O. Astafiev, Y. Nakamura, D.
V. Averin, and J. S. Tsai, Nature {\bf 421}, 823 (2003).A. J. Berkley,* H.
Xu, R. C. Ramos, M. A. Gubrud, F. W. Strauch,

P. R. Johnson, J. R. Anderson, A. J. Dragt, C. J. Lobb, F. C.
Wellstood.,Science, {\bf 300,}1548

\bibitem{CQED} \ \ P. Domokos, J. M. Raimond, M. Brune, and S. Haroche, Phys. Rev.
A {\bf 52}, 3554 (1995); T. Pellizzari, S. Gardiner, J. I. Cirac, P. Zoller, Phys.
Rev. Lett. {\bf 75}, 3788 (1995); L. You, X. X. Yi, and X. H. Su, Phys. Rev. A%
{\bf 67}, 032308 (2003); X. X. Yi, X. H. Su and L. You, Phys. Rev. Lett {\bf %
90}, 097902 (2003)

\bibitem{adiabatic elimination}S. B. Zheng and G. C. Guo, Phys. Rev.
Lett. {\bf 85}, 2392 (2000)

\bibitem{aa} A. Messiah  Quantum Mechanics vol 2 (Amsterdam: North-Holland)

\bibitem{BO} \ \ M. Born, R. Oppenheimer, Ann, Physik {\bf 84}, 457 (1930).

\bibitem{sun1} C. P. Sun, X. F. Liu, D. L. Zhou and S. X. Yu, Phys. Rev. A 
{\bf 63}, 062111 (2000); C. P. Sun, D. L. Zhou, S. Y. Yu and X. F. Liu, Eur.
Phys. D, {\bf 13}, 145(2001)

\bibitem{nano} Xue Ming Henry Huang, Christian A. Zorman, Mehran Mehregany, and M. L. Roukes, Nature (London) {\bf 421}, 496 (2003) 

\bibitem{sun2} C.P.Sun, M.L.Ge, Phys.Rev.D, 41, 1349, 1990.

\bibitem{Averin} D. V. Averin and C. Bruder, Phys. Rev. Lett. {\bf 91}, 057003(2003)

\bibitem{sun3} C. P. Sun, H. Zhan, X. F. Liu, Phys. Rev. {\bf 58}, 1810(1998)

\bibitem{Jx}   K. Molmer and A. Sorensen, Phys. Rev. Lett. {\bf 82}, 1835(1999)

\bibitem{zeng} B. Zeng, D. L. Zhou, C. P. Sun and L. You, private communication 

\end{references}
\end{document}